\documentclass[aps,prl,twocolumn,showpacs,superscriptaddress,amsmath,amssymb]{revtex4}

\usepackage{graphicx}
\usepackage{dcolumn}% Align table columns on decimal point
\usepackage{bm}% bold math
\usepackage{color}
\usepackage{txfonts}
\usepackage{comment}

\begin{document}
\title{Exploring Maps with Greedy Navigators}

\author{Sang Hoon Lee}
\email[Corresponding author.\\]{sanghoon.lee@physics.umu.se}
\affiliation{IceLab, Department of Physics, Ume{\aa} University, 901 87 Ume{\aa}, Sweden}

\author{Petter Holme}
\affiliation{IceLab, Department of Physics, Ume{\aa} University, 901 87 Ume{\aa}, Sweden}
\affiliation{Department of Energy Science, Sungkyunkwan University, Suwon 440--746, Korea}
\affiliation{Department of Sociology, Stockholm University, 106 91 Stockholm, Sweden}

\begin{abstract}
During the last decade of network research focusing on structural and dynamical properties of networks, the role of network users has been more or less underestimated from the bird's-eye view of global perspective. In this era of global positioning system equipped smartphones, however, user's ability to access local geometric information and find efficient pathways on networks plays a crucial role, rather than the globally optimal pathways. We present a simple greedy spatial navigation strategy as a probe to explore spatial networks. These greedy navigators use directional information in every move they take, without being trapped in a dead end based on their memory about previous routes. We suggest that the centralities measures have to be modified to incorporate the navigators' behavior, and present the intriguing effect of navigators' greediness where removing some edges may actually enhance the routing efficiency, which is reminiscent of Braess's paradox. In addition, using samples of road structures in large cities around the world, it is shown that the navigability measure we define reflects unique structural properties, which are not easy to predict from other topological characteristics. In this respect, we believe that our routing scheme significantly moves the routing problem on networks one step closer to reality, incorporating the inevitable incompleteness of navigators' information.
\end{abstract}

% PACS, the Physics and Astronomy % Classification Scheme.
%%89.40.-a Transportation
%%89.75.Fb Structures and organization in complex systems
%%89.75.-k Complex systems

%\keywords{}
%Use showkeys class option if keyword %display desired

\maketitle

A sociopsychological problem that has turned out to be especially suited to methods of physics is human mobility---how can we characterize, measure and explain the movement of people in their daily lives~\cite{CSong2010}? One piece of the human mobility puzzle is how to measure the navigability of cities and buildings. How successful we are at finding our way is a function both of our cognitive ability, how much relevant information we have, and also the spatial organization of our surroundings. The methods characterizing the spatial organization of cities and buildings can assume either that agents have complete information of the environment of their journey or that they essentially do random walks without any information. In the former category, there are measures like betweenness centrality~\cite{Wasserman,KIGoh2001}; in the other category, random walk centrality~\cite{JDNoh2004} or first-passage time~\cite{RednerBook}. The reality, of course, is somewhere in between---we always navigate with incomplete information~\cite{IncompleteInformation,Kleinberg2000,Boguna2008,SHLee2011}. This information could be better (if we have GPS devices or maps) or worse (going back to the cafeteria the second day at work in a big office building), but to understand the large-scale patterns of human movements and how it is influenced by the spatial organization, we need models and measures that incorporate navigation with incomplete information.

The key component of our approach is greedy navigators. These are agents in a spatial network~\cite{Barthelemy2011} that travel between a start $s$ and target point $t$. The agents have a sense of spatial orientation and a memory of where they have been. Briefly speaking, agents move in a direction as close as possible to the direction of the target. If they reach a cul-de-sac, they backtrack to the previous point where they have an untested choice of route. In this Letter, we use a discrete spatial network formalism, but it should be rather straightforward to extend the concepts to a continuous space. We stress that we do not try to model people's behavior exactly, but that the greedy navigators capture the relative magnitude of deviation from an optimal navigation caused by the underlying spatial structure. In other words, we surmise the greedy navigators fare worse in cities or buildings where humans would easily get lost or make unnecessary detours, than in those where it is easy to get around.

There are three main categories of cognitive processes in human navigation: the use of spatial cues, computational mechanisms, and spatial representations~\cite{wolbers,Thomas2007}. In our greedy spatial navigation (GSN) model, we assume the agents have a good, large-scale sense of the navigation, but no reliable real or cognitive map. Note that we focus on the spatial orientation~\cite{Arleo2001} rather than the geometric proximity~\cite{Kleinberg2000,Boguna2008,SHLee2011}, since the former is more comprehensible in the navigator's point of view inside the spatial structures based on the visual cue such as landmarks~\cite{wolbers}. In other words, it is more intuitive to think of as ``going to the road to the northern side because I know that the festival takes place in the northern part of the town'' than ``going to the road which will lead me to the closest point to the festival.'' Furthermore, the geometric-proximity-based strategies heavily depend on the choice of points (vertices or nodes in the language of graph or network), while the direction-based strategies are much more robust to the different choices of such points. Therefore, the latter seems to be a more reasonable choice, considering the fact that the spatial structures are continuous in reality.

\begin{figure}[b]
\includegraphics[width=0.9\columnwidth]{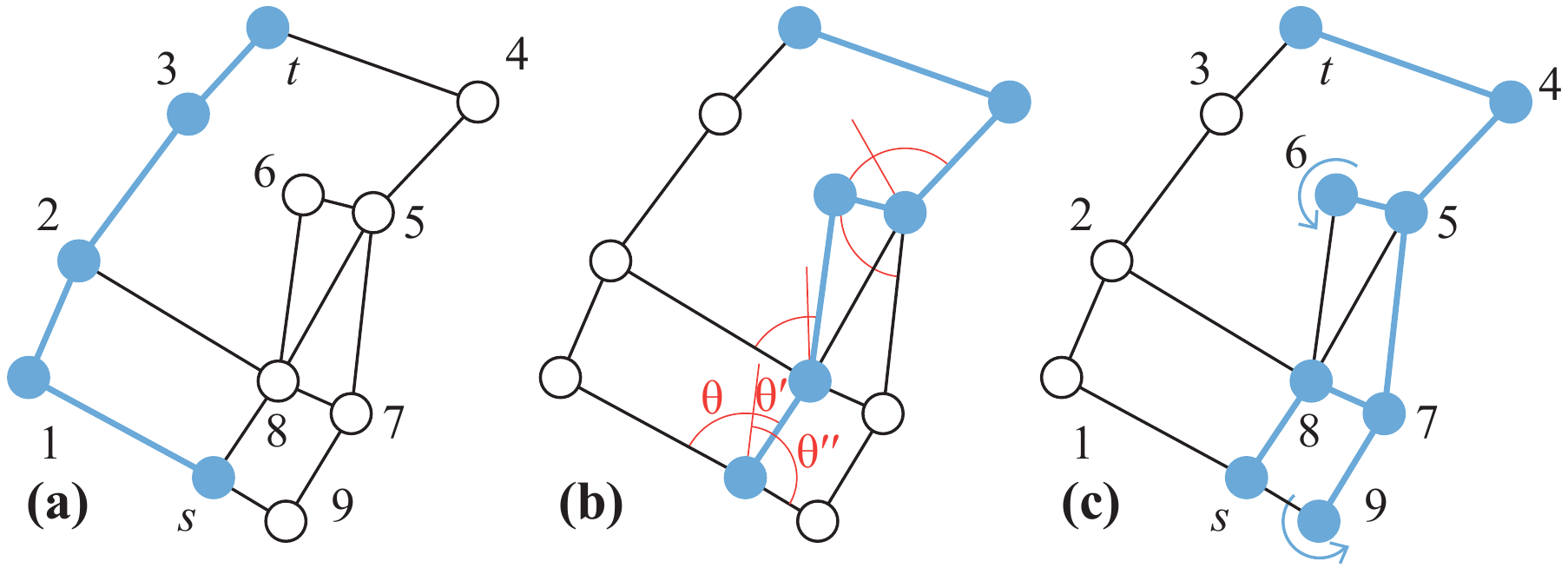}
\caption{Illustration of three different routing schemes to probe navigability-related structures of cities and buildings. (a) shows SPN from $s$ to $t$; (b) shows the route of GSN (the angles are those used in the algorithm); (c) illustrates a possible route of a random DFS navigation, where the backtracking steps occur at vertices $6$ and $9$.
}
\label{DFS_illustration}
\end{figure}

Translated to a spatial network language---where the network is represented as a graph of $N$ vertices at coordinates $\mathbf{r}_1,\dots,\mathbf{r}_N=(x_1,y_1),\dots ,(x_N,y_N)$ that are connected by $M$ edges---the greedy navigators act as follows. Assume an agent stands at a vertex $i$ and wants to travel to $t$. Let $\mathbf{v}_{i,j}=\mathbf{r}_j-\mathbf{r}_i$ be the vector between vertices $i$ and $j$ and $\theta_j$ be the angle between $\mathbf{v}_{i,t}$ and $\mathbf{v}_{i,j}$. The greedy navigator moves to the neighbor $j$ of $i$ that has not been visited before and has the smallest $\theta_j$. If all the neighbors of $i$ have been visited the navigator goes back to the vertex from which the navigator arrived to $i$, which is in contrast to the simple greedy navigation based on the geometric proximity that sometimes fails to reach the target due to the lack of such a backtracking process~\cite{Kleinberg2000,Boguna2008}. This procedure is repeated until $t$ is reached (which will happen eventually if $G$ is connected, or, more generally, if $i$ and $j$ belong to the same component). We contrast the greedy navigators with random navigators that move between the source and target in the same way as the greedy ones except that they go to a random neighbor instead of the one with minimal $\theta_j$. Essentially this is a random depth-first search (DFS)~\cite{DFS}. See Fig.~\ref{DFS_illustration} for an illustration.

How can we use greedy navigators to quantify the navigability of a map? Let $d_g$ ($d_r$) be the average distance in the eyes of the greedy (random) navigators, respectively. More precisely, we average, over all pairs of distinct vertices, the number of edges in the navigators' paths. $d$ is the average distance as usual [the average number of edges for the real shortest path navigation (SPN)]. $d$ is the lower bound of $d_g$ ($d_r$), which makes the greedy navigability $\nu=d/d_g$ (random navigability $\zeta = d/d_r$) a natural measure of greedy (random) navigability of the underlying spatial network, respectively. $\nu$ or $\zeta$ takes values in the interval $(0,1]$ where fewer detours means a higher value. The advantage of using the graph distance is that $\nu$ can be interpreted, roughly, as the fraction of correct choices of which road to take at an intersection~\cite{comment1}. We first measure the navigability of empirical data sets. Two of these maps are excerpts of the road networks of the Boston and New York City (NYC)~\cite{HYoun2007}. The roads of the excerpt are chosen to represent the major thoroughfares of the downtown areas. Other networks are railway networks from Europe~\cite{Kurant_railway}---a data set covering most of western continental Europe and the Swiss subnetwork of the former.

\begin{table}[b]
\caption{\label{navigability} Properties of five empirical data sets,
indicated by the performance of routing strategies.
See the text for the definition of symbols.
Null models for Boston and NYC roads are connected samples of
Erd\H{o}s-R{\'e}nyi random graphs~\cite{Erdos1959} with the same $N$ and $M$,
where the geographic layout to guide the GSN navigator is given
by the Kamada-Kawai algorithm~\cite{SHLee2011},
and the results averaged over $10^3$ samples are shown.}
\begin{ruledtabular}
\begin{tabular}{lccccccc}
network & $N$ & $M$ & $d_g$ & $d$ & $d_r$ & $\nu$ & $\zeta$ \\
\hline
Boston & 88 & 155 & 6.82 & 5.72 & 30.75 & 84\% & 19\% \\
Null model & & & 8.606(9) & 3.6758(1) & 23.20(1) & 43 \% & 16\% \\
%\hline
NYC & 125 & 217 & 8.27 & 6.79 & 44.39 & 82\% & 15\% \\
Null model & & & 11.72(2) & 4.0300(1) & 33.51(2) & 34 \% & 12 \% \\
%\hline
Switzerland & 1613 & 1680 & 145.14 & 46.56 & 769.68 & 32\% & 6 \% \\
%\hline
Europe & 4853 & 5765 & 143.69 & 50.87 & 2011.93 & 35\% & 3\% \\
%\hline
LCM (Fig.~\ref{fig:maze}) & 184 & 194 & 62.82 & 20.65 & 86.23 & 33\% & 24 \% \\
\end{tabular}
\end{ruledtabular}
\end{table}

\begin{figure}[t]
\includegraphics[width=0.8\columnwidth]{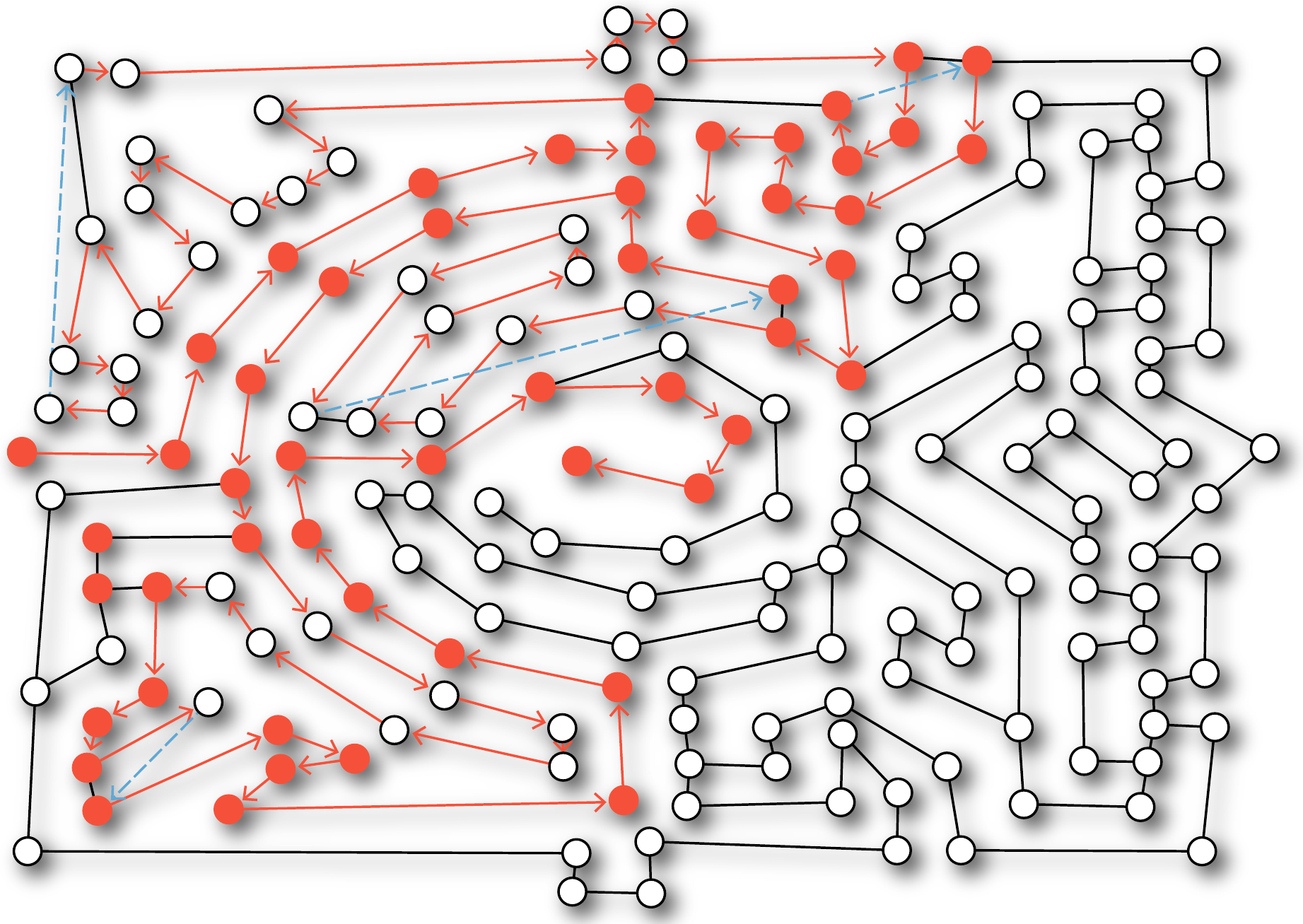}
\caption{An example of the GSN pathway from the entrance $\epsilon$ in the left to
the central target $\tau$ for LCM ($184$ vertices and $194$ edges)
is shown,
where the red arrows indicate the pathway with the backtracking
processes represented as blue dotted arrows and the SPN path
is shown in filled circles. The distances for this specific pair of the source and target with different
routing strategies (GSN, random DFS, and SPN, respectively) are
$d_g (\epsilon \to \tau) = 87$, compared to
$d_r (\epsilon \to \tau) = 134.13$ (averaged over $10^3$ trials with the standard error $1.067$),
and $d (\epsilon \to \tau) = 52$.
}
\label{fig:maze}
\end{figure}

The results for the navigability is shown in Table~\ref{navigability}.
For all the cases, $\nu$ is significantly larger than $\zeta$, which can be
intuitively understood since the real road or railway structures are designed by encoding geometric information useful to GSN.
More quantitatively, the real structures show a much better greedy navigability compared to a
network layout model for the visualization purpose as the null model~\cite{SHLee2011}, as we can check the cases of
Boston and NYC roads in Table~\ref{navigability} ($\nu / \zeta = 4.509$ for the real Boston roads and $2.696$
for the corresponding null model). One can also check that $\zeta$ strongly depends on the system size, which will be
discussed later in a systematic approach based on larger data sets.
As an example of the GSN, we show
the performance of GSN in case of an intentionally delusive layout---namely, a maze. Figure~\ref{fig:maze} shows an example of GSN
pathway on the graph representation of
Leeds Castle Maze (LCM)~\cite{LeedsCastleMaze} in England, starting from the entrance in the left to
the central target.

\begin{figure*}[t]
\includegraphics[width=0.85\textwidth]{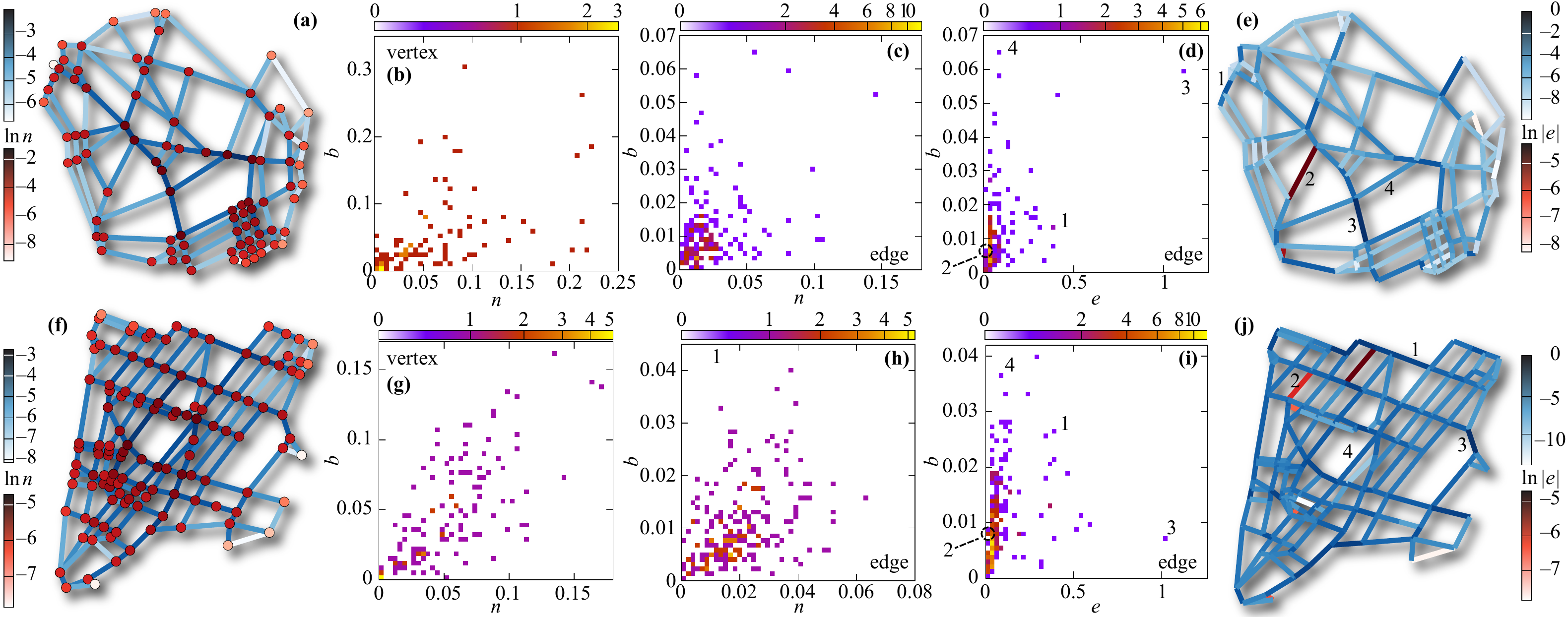}
\caption{Characteristics of vertices and edges of two empirical road maps from Boston (a)--(e) and NYC (f)--(j). We show scatter plots with color-coded density resolving the spatial structure with respect to vertices [(b) and (g)] and edges [(c), (d), (h), and (i)]. We show the relation between betweenness $b$ and navigator centrality $n$ [(b), (c), (g), and (h)] and the relation between navigator essentiality $e$ and betweenness $b$ [(d) and (i)]. The maps [(a) and (f)] show the natural logarithm of the navigator centrality for the vertices (the lower color bars) and edges (the upper color bars). [(e) and (j)] chart the natural logarithm of the absolute values of navigator essentiality [the upper (lower) color bars for normal (Braess) edges, respectively]. In (e) and (j) we also highlight some of the representative $e$ values from panels (d) and (i) respectively.}
\label{fig:empirical_examples}
\end{figure*}

Contrary to the conventional network centralities based either on the shortest path~\cite{Wasserman,KIGoh2001}
or on the random walk~\cite{JDNoh2004},
we define centralities based on our GSN and compare those with conventional ones in various cases.
First, the centrality considering the ``betweenness'' in the pathway, called
navigator centrality $n$ for vertex or edge $x$ is defined as
$n(x) = \sum_{i \ne j} \sigma_{ixj} / [N(N-1)]$,
where $\sigma_{ixj}=1$
if the GSN path from the source vertex $i$ and the target vertex $j$ goes
through the vertex or edge $x$ in the middle and $\sigma_{ixj}=0$
otherwise.
We present the Boston and NYC roads where the $n$ values are color coded, in Figs.~\ref{fig:empirical_examples}(a) and (f).
In all the networks including railway networks (not shown),
we have found that the correlation between $n$ and
conventional betweenness centrality $b$~\cite{Wasserman,KIGoh2001}
(using the shortest path only considering
topology) are larger for vertices than edges, as
shown in Figs.~\ref{fig:empirical_examples}(b), (c), (g), and (h). In other words, while the relative $b$ of vertices
is more or less similar whether it is defined from the SPN pathways
or from the GSN pathways, $b$ for the edges is different for those
two cases.

To further investigate the properties of edges in terms of GSN,
we introduce another centrality addressing the essentiality $e$ for edge $l$
defined as,
$e(l) = d_g [G \backslash \{l\}] - d_g [G]$
where $G \backslash \{l\}$ refers to the graph with the edge $l$ removed from $G$,
which quantifies the average number of additional steps necessary
as the effect from the absence of the edge $l$, naturally implying the
edge's importance for GSN. Note that $e(l)$ can be
negative somewhat counterintuitively, which implies that the removal of
the edge $l$ improves the navigability in terms of GSN.
The case $e(l) < 0$ is clearly reminiscent of
Braess's paradox~\cite{HYoun2007}; i.e., road closures
can sometimes reduce travel delays caused by the discrepancy between the
user-based optimum and system-level global optimum. Our example, therefore, illustrates an
interesting  phenomenon happening even to a single navigator that only comes from
the greedy navigation strategy.
Here we denote the edges with $e > 0$
as ``normal'' edges and the edges with $e < 0$ as ``Braess'' edges.
Figures~\ref{fig:empirical_examples}(e) and (j) show Boston and NYC roads
where $e$ values are color coded.
Again, as in case of $n$,
there is significant difference between $e$ and $b$. Therefore, we
conclude that the spatial structure of edges indeed acts as a crucial
substrate for greedy navigators.

In our example road structures, we observe that the detailed structure of networks really matters.
Figures~\ref{fig:empirical_examples}(d) and (i) clearly demonstrate
the diversity of road structures with the four representative roads
for each city. The examples clarify that relatively low correlations for
$e$ and $b$ stem from those roads whose importance is quite different for
GSN and SPN. Roads in the periphery show, expectedly, low
$e$ values, but there are some important exceptions in terms of GSN, e.g., road 1 in
Fig.~\ref{fig:empirical_examples}(e) with the large
$e$ value. In case of the Braess road 2 in Boston [Fig.~\ref{fig:empirical_examples}(e)],
we observe that its absence helps the
large volume of traffic from the upper left part to avoid entering
the central part to reach the lower right part, and induce to take more
efficient peripheral roads.
Of course, the external geographic factors such as rivers, tunnels, bridges,
and roads with various speed limits are also important in practice.
We take the simplest approach and assume the geographical context primarily gives a sense of direction for the navigation, and neglect other effects.
For future work it would be interesting to extend our work with other information into other navigability functions, e.g.,
Bureau of Public Roads (BPR) function~\cite{HYoun2007}. We also notice that
road 3 in Fig.~\ref{fig:empirical_examples}(e) with the largest $e$
value (and the second largest $b$ value) corresponds to the Harvard bridge across the Charles River, illustrating the
case of deducing the crucial infrastructure based solely on the geometric positions,
without explicit awareness of the river.

\begin{table}[t]
\caption{Coefficients for the multiple linear regression
$e = m_1 b + m_2 (\textrm{length}) +
m_3 c + m_4 (k_i k_j )$
for road networks,
with some measures defined on edges: $b$, the edge length, the distance $c$ from the midpoint of edges
to the centroid of vertices, and the product $k_i k_j$ of degrees of vertices attached to
edges.
The statistical significance codes
are
$< 0.05$, $< 0.01$, and $< 0.001$.}
\label{table2}
\begin{ruledtabular}
\begin{tabular}{lcc}
%\hline
%\hline
Road & Boston & NYC \\
\hline
$m_1$ & $6.902$\footnotemark[3] & $9.389$\footnotemark[3] \\
$m_2$ & ${-4.687 \times 10^{-5}}$\footnotemark[1] & ${-6.141 \times 10^{-5}}$\footnotemark[2] \\
$m_3$ & $-1.504 \times 10^{-6}$ & ${2.142 \times 10^{-5}}$\footnotemark[1] \\
$m_4$ & ${-8.817 \times 10^{-3}}$\footnotemark[2] & ${-5.653 \times 10^{-3}}$\footnotemark[1] \\
%\hline
\\
Multiple $R^2$ & $0.2508$ & $0.1917$ \\
$p$ value & $7.784 \times 10^{-9}$ & $3.395 \times 10^{-9}$ \\
%\hline
%\hline
\end{tabular}
\end{ruledtabular}
\footnotetext[1]{$< 0.05$.}
\footnotetext[2]{$< 0.01$.}
\footnotetext[3]{$< 0.001$.}
\end{table}

\begin{figure}[t]
\includegraphics[width=0.9\columnwidth]{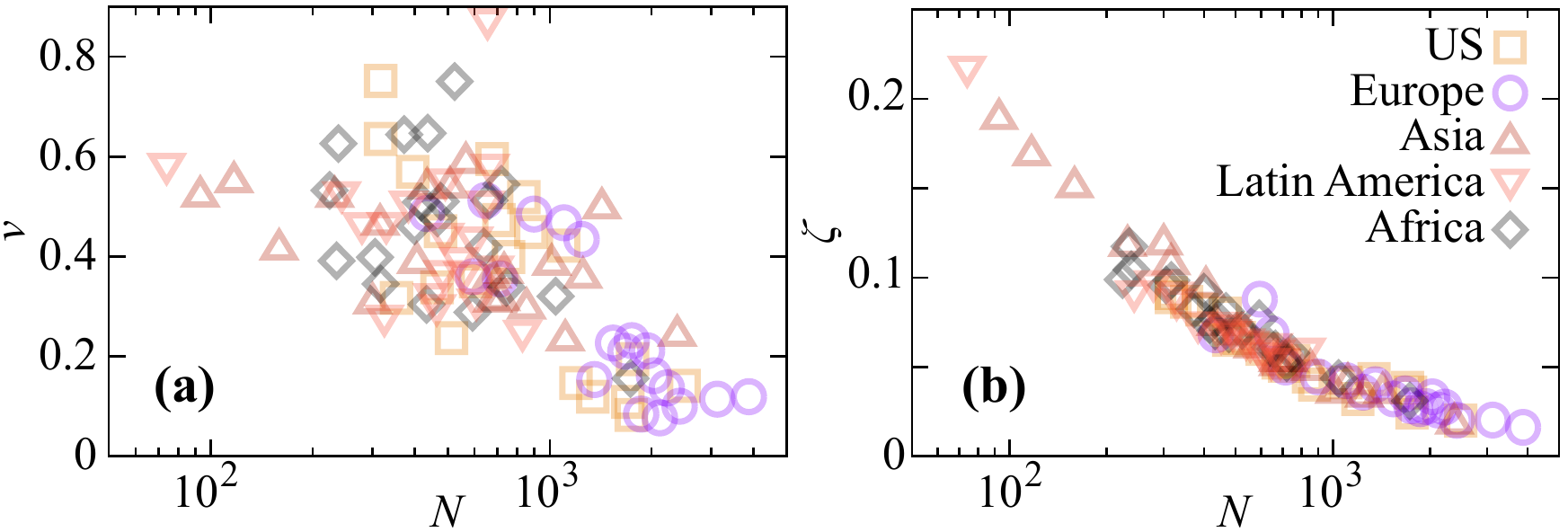}
\caption{Scatter plots for the $\nu$ (a) and
$\zeta$ (b) vs the number
of vertices $N$, for the $100$ large cities in the world.
}
\label{fig:nu_rDFS}
\end{figure}

The multiple linear regression results shown in Table~\ref{table2} demonstrate
that predicting $e$ values is not plausible from the linear
combination of those network and geometric measures, with low $R^2$ values.
From the same regression analysis on much larger Switzerland and European railways,
we observe even smaller $R^2$ values estimated by the $10^4$ sampled source-target pairs
for each removal of edge.
Therefore, $e$ or the Braessiness is a uniquely measured only by considering
this greedy behavior of navigators.
Finally, we investigate whether there is any correlation between
navigability and various socioeconomic indices. We selected the $20$ largest cities in
the United States (U.S.), Europe, Asia, Latin America, and
Africa, respectively ($100$ cities in total),
and used the \texttt{MERKAATOR} program~\cite{Merkaator} and extracted a representative
sample of each city (a square of $2 \textrm{km}$ sides)~\cite{road_network_dataset}. First,
we compared $\nu$ and $\zeta$ to the numbers of vertices $N$, as shown in Fig.~\ref{fig:nu_rDFS}.
There is a striking difference between those two cases, where
there is a clear scaling relationship between $\zeta$ and $N$ [Fig.~\ref{fig:nu_rDFS}(b)], meaning that the random navigation is statistically determined by the
system sizes. In contrast, the widely scattered points in Fig.~\ref{fig:nu_rDFS}(a)
strongly suggests that the numbers of vertices cannot predict
$\nu$ at all, in addition to the fact that purely topological
measures cannot predict $e$ in Table~\ref{table2}. In this respect, the $\nu$ obviously reflects
unique properties of different cities with vastly different developmental
histories.
We could not find such measures (or linear
combinations of them)---e.g., population density,
median resident income, fraction of public transit
commuters, etc.---showing statistically significant
correlations with the navigability.  Again this leads to the conclusion
that different cities have unique properties of navigability
independent of other socioeconomic factors. One example is
the correlation between
the navigability and the population change ratio of the $20$
cities in the U.S. defined as the ratio of the population change
between $1960$ and $2010$ to the population in $1960$~\cite{US_population}.
We observe a very weak negative correlation between $\nu$ and the
ratio [$R^2 = 0.09(3)]$---too weak perhaps for claiming a meaningful conclusion dependence.

In summary, we have introduced a new routing strategy incorporating greedy
movement and memory of navigators. This strategy, we believe, is a
minimal model considering the basic concept of human psychology for navigation,
namely, incomplete navigational information and the memory not to be lost.
From the results from real-world road and railway structures, we demonstrate
the important difference in terms of centralities for navigation
and the fact that there exists the celebrated Braess's paradox caused by the navigators'
behavior just equipped with this simple strategy. From the observation
of correlation profiles for centralities in road structures, we have shown
that the importance of each element heavily depends on the detailed layout of structures.
We have focused on the final efficiency of the routing processes in this work, but
the detailed process of GSN, e.g., the relative distance toward the target
during the routing process or the prevalence of backtracking related to the
structural properties of roads, can be worthwhile future work.
This type of tool---linking spatial cognition, the environment, and emergent navigational properties---can be helpful for urban planners and architects~\cite{hillier_Carlson2010}.

\begin{acknowledgments}
This research is supported by the Swedish Research Council and the WCU program through NRF Korea funded by MEST R31--2008--10029 (P.\,H.). The authors thank Vincent Blondel, Daniel Equercia, Veronica Ramenzoni, Bo S{\"o}derberg, and Hang-Hyun Jo for comments, and Hyejin Youn for help with data acquisition.
The computation was partially carried out using the cluster in CSSPL, KAIST.
\end{acknowledgments}


\begin{thebibliography}{00}
\bibitem{comment1}
Note that at a step the agent can either get one step closer or farther from the target or remain at the same distance. Assuming that the steps increasing the distance are rare in real networks the interpretation of $\nu$ as a correct-step rate is adequate.
\end{thebibliography}
\end{document}